# The Concurrent Language Aldwych

Matthew Huntbach

Department of Computer Science, Queen Mary and Westfield College
Mile End Road, London, UK, E1 4NS
`mmh@dcs.qmw.ac.uk`

**Abstract:** Aldwych is proposed as the foundation of a general purpose language for parallel applications. It works on a rule-based principle, and has aspects variously of concurrent functional, logic and object-oriented languages, yet it forms an integrated whole. It is intended to be applicable both for small-scale parallel programming, and for large-scale open systems.

## 1. Introduction

Actors [9], [1] and concurrent logic languages [21] have been proposed as the two paradigms most suitable as a foundation on which to design languages suitable for open systems [15]. The advantage offered by actors of the clarity of direct representation of objects is offset by the greater flexibility of direct representation of channels and the possibility of multiple input ports in concurrent logic languages [14]. Also actors offer many-to-one communication directly but one-to-many communication is awkward and must be managed by the programmer, while with concurrent logic languages it is the other way round.

In this paper we define a language which combines aspects of both Actors and the concurrent logic languages. Previous attempts to build actor languages on top of concurrent logic languages have, as noted by Kahn [14], tended to lose the advantages of the underlying logic language by modelling the actor principle too closely. We avoid this trap, combining the advantages stated above of both actors and concurrent logic languages.

## 2. A Language of Rewrite Rules

We start by describing a simple rewrite language [18] for manipulation of futures. The term "future" is used in parallel functional [8] and object-oriented [17] languages to refer to a value in a language which may be passed as an argument or incorporated in a structure simultaneously with being written to by another process. A future has exactly one writer but any number of readers, it may be bound to a tuple containing further futures. A reader of a future becomes a reader of any arguments of such a tuple, a writer becomes a writer to arguments of the tuple to which it has bound a future. Thus it is more limited than the single-assignment logic variable as that may have several writers and "back communication" [3]. Processes which depend on the value of a future suspend until they learn of its value.

We initiate a process of class $p$ by $p(u_1, ..., u_m) \rightarrow (v_1, ..., v_n)$, the process is a writer to each future $v_i$ and a reader to each $u_i$. An ensemble of processes is created by a list of such initiations, where a future occurs exactly once in a $v_i$ position and one or more times in a $u_i$ position. A class is described by a header taking the form $\#p(x_1, ..., x_m) \rightarrow (y_1, ..., y_n)$ and a set of rules.

A rewrite rule has two parts, the left and right hand sides (lhs and rhs), the lhs representing the conditions needed for a rewrite to take place, the rhs giving the rewrite. The lhs consists of a list of equality statements, `<Future>=<Tuple>`, where `<Future>` is any input future named in the header which does not occur in another equality statement in the same lhs, and `<Tuple>` a tuple or constant, whose arguments may be further tuples or constants or new futures. The rhs consists of a list of process creations in which the input futures are new futures or any future from the header or any future from tuples on the lhs, plus mandatory equality statements `<Future>=<Tuple>` for every output future from the header not used in a process creation. Any future which occurs only on the rhs must occur in exactly one output position and one or more input positions, any future from a tuple on the lhs must occur at least once in a process creation or a rhs equality statement. The equalities on the rhs are the mechanism by which futures are written to, and have an alternative form `<Future>←<Future>` which binds an output future to an input future. The restrictions ensure that all futures have exactly one writer. As an example:

```
#p(u,v)→(x,y)
{
 u=f(w)         ||  r(u,v,w)→(y,x);
 u=g(w)         ||  q(v,w)→x, y=g(v);
 u=h(w), v=a    ||  x←w, y=b;
 v=e            ||  x=k(u,z), s(u)→(z,y)
}
```

is the syntax used to indicate that a `p` process waits on two futures it calls `u` and `v`, and is a writer for two futures it calls `x` and `y`. If `u` becomes bound to `f(w)`, where `w` is another future, it creates an `r` process taking `u`, `v` and `w` as inputs (`u` is at this point necessarily bound to `f(w)`, but may still be used as input), and writing to `x` and `y`. If `u` becomes bound to `g(w)` it creates a `q` process taking `v` and `w` as inputs and writing to `x`, while `y` is bound to the tuple `g(v)`. If `u` becomes bound to `h(w)` and `v` becomes bound to the constant `a`, the constant `b` is written to `y` while `x` becomes bound to `w` (if `w` is not itself bound at this point, any reader of `x` will become a reader for `w`). If `v` becomes bound to `e`, `x` is bound to the tuple `k(u,z)` (where `z` is a new future) while the process `s` with argument `u` writes to `z` and `y`. Note that it is possible to determine whether something is a future or a constant by context, thus `x=w` in the place of `x←w` would cause `x` to be bound to the constant `w`.

If more than one rule is applicable, for example if `v` is bound to `e` and `u` to `f(w)` in a `p` process, either may be used. A process may react using one rule without waiting to see if another becomes applicable later, for example if `v` is bound to `e` but `u` not yet bound. Rewriting is a commitment, there is no Prolog-like backtracking. If rewriting is dependent on more than one future value, as in the third rule above, no commitment to any rule is made before all necessary variables are bound. For example, if `v` becomes bound to `a`, `p` is not committed to wait for `u` to become bound to `h(w)`, it will rewrite according to its first rule if `u` becomes bound to `f(w)`. Any number of rewrites may take place simultaneously if the system is implemented on a parallel architecture.

## 3. Rewrite Rules to Processes

So far our processes disappear as soon as they rewrite. Long-lasting processes are created with a variant form of rewrite rule, indicated by using a single bar to separate the lhs and rhs. In this case there is an implicit but unwritten recursive call. By default, the arguments to the unwritten recursive call are those of the original process. A statement `<Future>=<Tuple>` on the rhs, where `<Future>` is an input future in the header, sets the input argument of the recursive call in the position of `<Future>` in the header to `<Tuple>`. If an output future is not written to on the rhs, it takes its value from the equivalent output of the recursive call. If an output future is written to, either with an equality statement or by putting it as an output argument to a new process, any other use of that future is taken to refer to the output in the equivalent position in the recursive call. Thus,

```
#p(u,v)→(x,y)
{
 u=f(w)  |  u←w,  q(u,w,x)→x;
 u=g(w)  |  v=a,  r(v,x,w)→u;
 v=e    ||  x←u,  y=b
}
```

is a shorthand for:

```
#p(u,v)→(x,y)
{
 u=f(w)  ||  q(u,w,x1)→x, p(w,v)→(x1,y);
 u=g(w)  ||  v1=a, r(v,x,w)→u1, p(u1,v1)→(x,y);
 v=e    ||  x←u, y=b
}
```

In effect, the arguments to a process may be considered its state, with any changes in the recursive call representing a change of state. We now have something similar to the Actors concept. An actor reacts on receiving a message by doing three actions: *creating* new actors, *sending* messages to other actors, and *becoming* a new actor. In our language, a process reacts on receiving a future binding by *creating* new processes, *binding* other futures, and *becoming* its recursive call.

## 4. Futures to Channels

Futures seem to provide once-only communication, rather than the extended communication across channels of notations like CSP [10]. However a value which is constructed piece-by-piece can be regarded as a stream of information rather than seen in its entirety. A future may be treated as a stream or channel communicating a value if it is bound to a tuple with two arguments, one the value, the second a further future representing the remainder of the stream or the continuation of the channel. For example, the following represents a simple process which has a single input channel and a single output channel. It sends out a `b` when it receives an `a` and a `c` when it receives a `d`, and terminates when it reaches the end of the input stream, indicated by it being bound to the constant `end`, it then similarly ends its output stream.

```
#p(in)→out
{
 in=f(x,y), x=a  |  z=b, in←y, out=f(z,out);
 in=f(x,y), x=c  |  z=d, in←y, out=f(z,out);
 in=end         ||  out=end
}
```

The syntax, however, is awkward, so we introduce further notation, where

```
#p(in)→out
{
 in.a  | out.b;
 in.c  | out.d;
 in$   || out$
}
```

is the equivalent to the above, except a standard cons operator (expressible using infix `:`) is used in the place of `f`, and a standard nil constant (expressible as `$`) is used for `end`. In general, `x.<Tuple>` on the lhs of a rule, where `x` is any input future name, is a shorthand for `x=<Tuple>:x'` combined with a replacement of all occurrences of `x` on the rhs by the new name `x'`. We can send any tuple not just constants as in the example above. On the rhs of a rule, `x.<Tuple>` is equivalent to `x=<Tuple>:x`, where we already have any input use of `x` referring to the future in the `x` position in the recursive call. `x$` is equivalent to `x=$` on both lhs and rhs.

Note that since `in.a` inputs the constant `a`, we need another notation to input a non-defined future, and similarly to output one. We use `x?y` on the lhs to mean `x=y:x`, where `y` is a new future, and `x^y` to output `y` on output channel `x`. So, for example, an indeterminate merger of two channels into one is given by:

```
#merge(in1,in2)→out
{
 in1?m  | out^m;
 in2?m  | out^m;
 in1$   || out←in2;
 in2$   || out←in1
}
```

Input and output of multiple messages is allowed, so `x?a.b` on the lhs matches against the input of any value (which may then be referred to as `a`) followed by constant `b` on channel `x`, while `y.b^c.d(a)` on the rhs indicates sending the constant `b`, followed by the future `c` followed by the tuple `d(a)` (where `a` is a future) on channel `y`. There is a lookahead mechanism, enabling a value in a channel to be viewed but not consumed, so `x?a/.b` on the lhs matches only when the constant `b` is the second value on channel `x`, but as it is not consumed the `b` becomes the first item on channel `x` in the recursive call.

## 5. Representation in a Concurrent Logic Language

The language proposed so far is actually a variant syntax of a restricted form of a concurrent logic language. Each rule is equivalent to a clause, replacing the pattern matching of standard logic programming with named argument matching. So the equivalent of our first program is:

```
p(f(W),V,X,Y) :- r(U,V,W,Y,X).

p(g(W),V,X,Y) :- q(V,W,X), y=g(V).

p(h(W),a,X,Y) :- X=W, Y=b.

p(U,e,X,Y) :- X=k(U,Z), s(U,Z,Y).
```

Restricting variables to have a strict input/output mode with a single writer just formalises a practice which is generally kept to by programmers in the concurrent logic languages. Enforcing this restriction through syntax aids the production of bug-free programs. The lack of back-communication is however a serious problem since this technique is commonly used in concurrent logic programming. We shall show how this problem is overcome in the next section. Note we do not require the convention, inherited from Edinburgh Prolog, of distinguishing between variables and constants by initial capital and small letters, since we make the distinction by context.

A major aim of our syntax is to remove the redundancy inherent in the attempt of the concurrent logic languages to hold on to the logic-like syntax of Prolog. We develop from the view of concurrent logic languages as variants of the object-oriented paradigm [22], [12] by introducing this syntax tailored to that view. The ability to translate to a concurrent logic language, which we shall keep for all additional features, gives both a semantics for our language and a means of easy implementation since efficient implementations of concurrent logic languages exist [19].

Our work has some aims similar to that of Foster and Taylor [4] who set out to promote a concurrent logic language, which they called "Strand" from "stream AND-parallelism", without placing any emphasis on its logic background. Our language is called "Aldwych", since it turns into the language Strand (or a similar one) and the street "Aldwych" turns into the street "Strand" on the London street-map. Strand was developed further into a notation called PCN [5], which like Aldwych makes use of the single-assignment logic variable but replaces pattern-matching in a logic-style clause by named argument matching. However, PCN kept Prolog's lack of syntactic distinction between input and output variables, which we regard as a major limiting factor on its development. Like PCN, Aldwych could be used as a coordination language [6] so long as foreign language components are given a suitable interface.

Simple conditional statements may be added to the lhs of Aldwych rules translating directly to the guards of flat concurrent logic languages. For example, an ordered merge of two ordered streams is given by:

```
#ordmerge(in1,in2)→out
{
 in1?m1, in2/?m2, m2>=m1 │ out^m1;
 in1/?m1, in2?m2, m1>=m2 │ out^m2;
 in1$ ││ out←in2;
 in2$ ││ out←in1
}
```

As a further removal of redundancy, we shall allow functional-like embedded calls on the rhs of rules. For example, `p(q(x))→y` will be interpreted as `q(x)→z,p(z)→y`. This aids clarity by reducing the name space (no need for new variable `z`) and the need to match output in one place with input in another through an extra variable. Any embedded call will be assumed to have just one output argument.

## 6. Introducing Objects

In a concurrent logic language, several objects may access one object if each has a channel to which it writes, and these channels are merged to form a single channel which the object being accessed reads. Each writing object may regard its channel as a handle on the reading object. Thus:

```
p(pargs)→s1, q(qargs)→s2, r(rargs)→s3, s(s0,sargs),
merge(s1,s2)→s4, merge(s4,s3)→s0
```

Here, the `p`, `q` and `r` objects have `s1`, `s2` and `s3` as their respective handles on the `s` object, with the `s` object receiving a single stream of messages from them in `s0`. To simplify this, we add handles to Aldwych as a different type of value from futures. We distinguish them by using names beginning with capital letters. References to a handle are converted into channels with mergers added as appropriate for implementation. The above could be given as:

```
p(pargs)→S, q(qargs)→S, r(rargs)→S, s(S,sargs)
```

but this is counter-intuitive. `S` should be viewed as an argument and hence an input of the `p`, `q`, and `r` objects. The handle of an object is better seen as an output from its initiator. So we write the system of a `p`, `q` and `r` object having common access to an `s` object as an acquaintance as:

```
p(S,pargs), q(S,qargs), r(S,rargs), s(sargs)→S
```

This is converted into the form with merged channels automatically for implementation. Syntactically, handle variables are treated like future variables, for instance as above they must be in exactly one output position, but may be in any number of input positions; they may be assigned to each other (though handles may not be assigned to futures and vice versa). So:

```
#p(u,v,S,T)→H
{
 u=f(w)  |  q(w,S)→u,  r(v,w)→T;
 u=g(w)  |  r(v,w)→H,  s(H,T)→(u,v);
 u=h(R)  |  S←R,  q(x,S)→u,  s(R,T)→(x,v)
}
```

becomes, on making the recursion explicit:

```
#p(u,v,S,T)→H
{
 u=f(w)  ||  q(w,S)→u1,  r(v,w)→T1,  p(u1,v,S,T1)→H;
 u=g(w)  ||  r(v,w)→H,  s(H1,T)→(u1,v1),  p(u1,v1,S,T)→H1;
 u=h(R)  ||  q(x,S)→u1,  s(R,T)→(x,v),  p(u1,v,R,T)→H
}
```

and then with the conversion of handles to channels and the addition of the merges:

```
#p(u,v,h)→(s,t)
{
 u=f(w)  ||    q(w)→(s1,u1), r(v,w,t1), p(u1,v,h)→(s2,t1),
               merge(s1,s2)→s, t$;
 u=g(w)  ||    r(v,w,h), s()→(h1,t1,u1,v1),
               p(u1,v1,h1)→(s,t2), merge(t1,t2)→t;
 u=h(r)  ||    q(x,s)→u1, s()→(r1,t1,x,v),
               p(u1,v,h)→(r2,t2),
               merge(r1,r2)→r, merge(t1,t2)→t, s$
}
```

Note how an input object which is unused converts to an output channel which is closed by assigning the nil constant, `$`, to it.

The third rule above causes us a problem. We wanted to deal with a case where an input tuple contained a handle, but the reversal of input/output polarity in the conversion of handles to channels would require back communication (the `r` in `h(r)` should be used for output) which we have ruled out. To prevent this we do not allow futures to be assigned tuples which contain handles or to be used as channels to send messages containing handles, so this is invalid Aldwych.

However, note how the rule that handles, like futures, have only one writer means they convert to channels with only one reader. This means we *can* allow them to be used for back-communication. The objection to back-communication was that the one-to-many communication meant messages intended for back-communication could become duplicated and thus the back-communication slots could have multiple writers. As this is not the case with channels derived from objects, we allow handles to be used to send messages which do have reply slots and which have arguments, both input and/or reply, which are themselves handles. As an alternative to a collection of merge processes, object variables could be implemented using the port mechanism described by Janson et al [13], which would be more efficient. Since this is operationally equivalent to a collection of merge processes, it keeps the concurrent logic semantics.

The syntax for reading and writing message to and from handles is the same as that for reading one from a channel, but messages are read from an output handle. Messages take the form `<Name>(<Input arguments>)→<Output Arguments>`, with argument names beginning with capital letters if they are intended to be handles, and the → omitted if there are no output arguments. All output arguments on the lhs of a rule must be given a value on the rhs by occurring exactly once in an output position. Messages may be sent on both input and output handles, the latter being equivalent to "sending a message to self" in standard object-oriented terminology, working by joining the message to the front of the incoming stream that forms the output handle. Thus:

```
#p(a,b,Q)→H
{
 H.f(x)→(y,z)  | q(a,b)→y, r(x,Q)→z;
 H.g(c,R)→S    | p(q(c,b),b,R)→S, b←c;
 H.h(c)→d      | Q.m(c,t(a,Q))→d;
 H.r(R)        | Q←R, s(b,Q,R);
 H.s(c)→S      | H.g(c,Q)→S;
 H$            || Q.bye(a)
}
```

defines a `p` process which has two input futures and one input object and outputs one handle. It reacts to five different sorts of messages received on the handle it outputs:

`f`: has one input future and two output futures; `p` computes these outputs using `q` and `r` processes.

`g`: has one input future, one input object and one output object; `p` sets up the output object by setting up another `p` object (with its `a` argument produced by a `q` process), and replaces its own `b` value by the value input in the message.

`h`: has one input future and one output future; `p` sends an `m` message to its own input object to obtain the output value, with an extra value produced by a new `t` process.

`r`: has just an input object; `p` changes its own input object to the new input object, and sets up a new `s` process (note though this process has no output it can affect the rest of the system by sending messages to its object inputs).

`s`: has an input future and an output object; `p` gives a value to the output object by sending itself a `g` message.

The final rule may be considered as providing a destructor function. It tells the `p` process what to do when there are no references anywhere to its output handle, in this case send a final `bye` message containing its input values to its input object.

Making the recursion explicit, converting the objects to tuples (including syntactically incorrect back-communication arguments to channel messages, indicated by italics), and expanding the embedded calls gives the following as the equivalent to the above:

```
#p(a,b,h)→q
{
 h=f(x,y,z):h1    ||  q(a,b)→y, r(x)→(q1,z), p(a,b,h1)→q2,
                      merge(q1,q2)→q;
 h=g(c,s,r):h1    ||  p(d,b,s)→r, p(a,c,h1)→q, q(c,b)→d;
 h=h(c,d):h1      ||  q=m(c,e,d):q3, p(a,b,h1)→q1,
                      t(a)→(q2,e), merge(q1,q2)→q3;
 h=r(r):h1        ||  s(b)→(q,r1), p(a,b,h1)→r2,
                      merge(r1,r2)→r;
 h=s(c,s):h1      ||  h2=g(c,s,q1):h1, p(a,b,h2)→q2,
                      merge(q1,q2)→q;
 h=$              ||  q=bye(a):$
}
```

Messages may be received on handles using `?`, and sent using `^`, with the rule that any message received using `?` must be referred to again exactly once in a message send using `^`, and no where else. This is because such a message is not a normal tuple as it may contain object arguments and back-communication arguments. Lookahead is permitted on object messages as well as channel messages, though an object message which is being looked at but not consumed has its output arguments treated as additional inputs, since they are only written to when consumed.

We allow arguments to tuples and messages and ← assignments on the rhs of rules to be expressions built up using the arithmetic and a few other infix. Values in expressions of the form `<Name>(<Args>)` are replaced by a new variable name, say `X`, with the call `<Name>(<Args>)→X` added to the statement, while `<Handle>.<Name>(<Args>)` is replaced by `X` with the object message send `<Handle>.<Name>(<Args>)→X` added. The `^` notation used to send messages on channels is extended so that if its right-hand argument is an expression the value of that expression is sent on the channel. It may not however be used if the left-hand argument is an object name and the right-hand argument anything but a variable name representing a previously read object message.

The form:

`S.p(a)-|> <Value>, …`

where `−`, `|` and `>` are separate symbols of the language, is equivalent to

`S.p(a)→x | x←<Value>, …`

The p(a)- message pattern may occur anywhere on a lhs, so long as it is the only anonymous return on that lhs. The statement > <Value> writes the value to the anonymous return.

Message patterns may be given without being attached to a particular output handle, in which case it is taken to be a shorthand for rules attaching them to every handle, so

```
#p(sum,count)→(Priv,Ord)
{
 inc(x)      | sum←sum+x;
 Priv.dec(x) | sum←sum-x;
 balance   -|> sum, count←count+1
}
```

is a shorter form of:

```
#p(sum,count)→(Priv,Ord)
{
 Priv.inc(x)      | sum←sum+x;
 Ord.inc(x)       | sum←sum+x;
 Priv.dec(x)      | sum←sum-x;
 Priv.balance→r   | r←sum, count←count+1;
 Ord.balance→r    | r←sum, count←count+1;
 Priv$, Ord$      ||
}
```

Note the ability to have more than one output handle allows different classes of access to an object. In this example, the Priv handle enables dec messages to be sent to a p object, whereas the Ord handle does not.

## 7. Cells, Pointers and Functional Programming

Objects may be linked together with handles in a way similar to the linking of cells with pointers in standard imperative programming data structures. So:

```
#empty()→Self
{
 isempty-|>=true;
 cons(X)-|>list(X,Self)
}

#list(Hd,Tl)→Self
{
 head-|>Hd;
 tail-|>Tl;
 isempty-|>=false;
 cons(X)-|>list(X,Self)
}
```

provides lists of objects using the linked list rule. If L represents a list, then the head of the tail of the tail of the list (assuming its length is known to be more than two) is obtained by ((L.tail).tail).head, while a list of three objects, A, B and C is obtained by ((empty().cons(A)).cons(B)).cons(C). Note that we allow

`f(<Args>).<Message>` to be the equivalent of `f(<Args>)`→X, X.<Message>, and `(<Expression>).<Message>` equivalent to X.<Message> where <Expression> evaluates to X.

The syntax `L~tail~tail~head` and `empty()~cons(A)~cons(B)~cons(C)` is provided as an alternative to the above, where in general, `L~f(<Args>)` is a value which, as with `L.f(<Args>)` before, evaluates to X, where `L.f(<Args>)`→X is added to the rhs of a rule. However, in this context any further message sent is taken to be sent not to L, but to the reply from `f(<Args>)`, so while the value `L.f(<Args>).g(<Args>)` is equivalent to X where L is sent first `f(<Args>)` then `g(<Args>)`→X, `L~f(<Args>)~g(<Args>)` is equivalent to Y where L is sent `f(<Args>)`→X, then X is sent `g(<Args>)`→Y.

Juxtaposition of two values is also used as a syntax for message sending, with the distinction between juxtaposition and ~ roughly equivalent to that between using ^ and using . for sending messages, that is it sends a message which is the result of an evaluation. The message sent is an empty name message with one reply, in other words it matches the lhs pattern `(<Args>)`→R. So, for example, d ← (f(a) P.m(b) g(c)) is equivalent to f(a)→X, P.m(b)→Y, X.(Y)→Z, g(c)→W, Z.(W)→d. Juxtaposition is left-associative, which means there is a distinction between the expression `F(<Expression>)` and the expression `F (<Expression>)` (with a space between the F and the bracket). The former treats F as a process name, taking one input argument the value of <Expression> and having one output argument which is returned as the value of the expression. The latter treats it as a variable representing an object to which the value of <Expression> is sent as the input argument to a nil-name message, the output argument being the value of the expression. The effect is to enable functional programming in the higher order style [2] to be embedded within Aldwych.

A function is an entity which when presented with a value returns a result, and in the higher order style such an entity is itself a first class citizen of the language. An object which is restricted to taking nil-messages and which does not change its internal state at any stage may be regarded as a function. An important aspect of functional programming is the idea of currying where a function may be supplied with a partial number of its arguments and return a function specialised to those arguments. Given a function f which takes n arguments, f $a_1$ $a_2$ … $a_{n-1}$ returns a function f' which when supplied with the final argument $a_n$ gives the same result as f when given all n arguments $a_1$ … $a_n$. A similar argument applies to f', and so on, hence functions need only take one argument supplied by juxtaposition.

In Aldwych, a header of the form:

`#p($a_1$,…,$a_k$) [$a_{k+1}$ … $a_{k+n}$]`→(`$v_1$,…,$v_m$`)

followed by a set of rules suitable for a header `#p($a_1$,…,$a_{k+n}$)`→(`$v_1$,…,$v_m$`) translates to a set of process declarations, starting with:

```
#p(a_1,…,a_k)→Self
{
 Self.(Val)→Return | p_1(a_1,…,a_k,Val)→Return;
 Self$ ||
}
```

which could be written in the briefer style introduced above:

```
#p(a₁,…,aₖ)→Self
{
  (Val)-|> p₁(a₁,…,aₖ,Val)
}
```

followed by (in the briefer style again):

```
#p₁(a₁,…,aₖ,aₖ₊₁)→Self
{
  (Val)-|> p₂(a₁,…,aₖ,aₖ₊₁,Val)
}
```

and so on, ending with with one with header $\#p_n(a_1,…,a_{k+n}) \to (v_1,…,v_m)$ and the original body. So if P is the output of $p_i(a_1,…,a_{k+i})$, then P X will be Y where the message (X)→Y is sent to P, which is the output of $p_{i+1}(a_1,…,a_{k+i},X)$, which is P specialised with the next argument set to X, and may be used as a first class object. P may be thought of as a higher-order function, but it can also be seen as an object-creating object, equivalent to a class in pure object-oriented languages like Smalltalk [7], where classes are themselves objects.

## 8. Initialisation, Default Values and Delegation

Many concurrent logic languages have an "otherwise" construct which separates clauses. Computation may only commit to a clause following the "otherwise" if variables have been sufficiently bound to rule out the possibility of committing to any clauses before it. We use the colon character as Aldwych's "otherwise", thus a procedure to return the maximum of two inputs is:

```
#max(a,b)<
{
 a>b ||> a;
:
     ||> b
}
```

Here the otherwise gives us a form of if-then-else. This also shows an extension to the anonymous return (< is used for anonymous return of a value in a header), allowing the return value for a whole process to be anonymous, giving a more functional appearance to the code. The symbol < may be used as a value elsewhere if it is necessary to use the anonymous output of a recursive call as input.

Delegation [17] has been recommended as an alternative to inheritance in a concurrent object-oriented system. The idea is that one of the arguments to an object is another object termed a "proxy", any messages an object cannot handle are passed to its proxy. Below is an example of its use to define the class of royal elephants, which are like ordinary elephants in every way except their colour is white:

```
#RoyalElephant(Proxy)~
{
 colour-|>=white;
:
 ?m | Proxy^m
}
```

The ~ indicates an anonymous handle return, the above could be written more fully

```
#RoyalElephant(Proxy)→Self
{
 Self.colour→return | return=white;
:
 Self?m | Proxy^m
}
```

A ~ as a value in an expression is used to mean `Self` as above, similar to < as above, while ~<Mess> is equivalent to `Self.<Mess>`. The delegation effect depends on royal elephant objects being declared with an appropriate proxy. We introduce a syntax allowing processes to have local variables which are initialised when the processes are created. In this case, a royal elephant class is declared by:

```
#RoyalElephant(Args)~
= Elephant(Args)→Proxy, <(Proxy)
{
 colour-|>=white;
:
 ?m | Proxy^m
}
```

In general, an = following a header indicates there is some initialisation code, this code must write values to all local input arguments, and use all local output arguments. An argument may be local to just the initialisation code provided it has exactly one writer and at least one reader within that code. The < indicates the declaration of local arguments, which are treated in the rules just like other arguments to the object. In effect, any call `RoyalElephant(Args)→E` is transformed to one `Elephant(Args)→Proxy, RoyalElephant'(Args,Proxy)→E`, where `RoyalElephant'` has the rules declared for `RoyalElephant`. It is assumed that `Args` hold the arguments needed for any elephant. So each royal elephant object has its own elephant object which acts as its proxy.

An alternative way of declaring this is:

```
#RoyalElephant(Args,Proxy←Elephant(Args))~
{
 colour-|>=white;
:
 ?m | Proxy^m
}
```

Here `Proxy` is an argument with a default value. In general, a default value argument is declared in a header by following it with ←<Expression>, where <Expression> is any expression using the arguments of the header as input. It has a similar effect to a local variable, but the default may be overridden when an object is created, thus the call

RoyalElephant(Args)→E works as before, but RoyalElephant(Args,Proxy←FunnyElephant(Args))→E creates a royal elephant object whose proxy is a funny elephant object, in other words it is equivalent to FunnyElephant(Args)→Proxy, RoyalElephant'(Args,Proxy)→E.

Note that == is used for initialisation where there is no body, in effect declaring a macro, for example (using anonymous output, so <, ==, and > are separate symbols):

```
#square(x) <==> x*x;
```

This can also be seen as a way of avoiding "single rule" processes like:

```
#square(x)<
{
  ||>x*x
}
```

## 9. More Control

The next addition to the language deals with the fact that programs in concurrent logic languages tend to have a tangled structure of mutually recursive procedures. We allow anonymous embedded procedures, with their bodies written at the end of the rhs of the rule where they are called. This reduces the name space of procedures, and takes away the goto-like jumping between the mutually recursive procedures. In general, an embedded procedure is declared by adding a set of rules, with the same syntax as top-level rules, at the end of the rule which calls it. The embedded procedure has the same header (including local arguments) as the calling procedure with the addition of the anonymous return for the rule if there is one and it has not been given a value in the rule. A call to it is made in the place of the recursive call, with the arguments the recursive call would have had. A double bar on an embedded procedure rule indicates that recursion reverts to the embedding procedure, a triple bar means that there is no recursion. In general, given there is no limit on the level of embedding, the number of bars is one greater than the number of embeddings which are exited.

So, the following declares a process with one input and one output channel which reads a series of constants, deleting those which are enclosed by the constants `stop` and `start`:

```
#delbetween(in)→out
{
 in.stop | {
           in.start || ;
          :
           in?m      | ;
           in$       ||| out$
          }
:
 in?m | out^m;
 in$  || out$
}
```

Using mutual recursion would make the above:

```
#delbetween(in)→out
{
 in.stop || delbetween1(in)→out;
:
 in?m   | out^m;
 in$    || out$
}
```

```
#delbetween1(in)→out
{
 in.start || delbetween(in)→out;
:
 in?m     | ;
 in$      || out$
}
```

Embedded procedures may have local variables which are declared in a similar way to local variables for top-level procedures, so, for example, the following is similar to the above but outputs a separate stream which sends a list of the number of characters deleted on after each occurrence of stop:

```
#delcountbetween(in)→(out,count)
{
 in.stop | num←0,<(num)
           {
            in.start ||  count^num;
           :
            in?m     |   num←num+1;
            in$      ||| count^num$, out$
           }
:
 in?m | out^m;
 in$  || out$, count$
}
```

Here the equivalent is:

```
#delcountbetween(in)→(out,count)
{
 in.stop || delcountbetween1(in,0)→(out,count);
:
 in?m | out^m;
 in$  || out$, count$
}
```

```
#delcountbetween1(in,num)→(out,count)
{
 in.start || count^num, delcountbetween(in)→(out,count);
:
 in?m     | num←num+1;
 in$      || count^num$, out$
}
```

Any output local variable must have an explicit writer in any rule whose number of bars is such that the variable is no longer in scope and hence not an argument to the mutually recursive call.

A variant of the embedded procedure gives an if-then-else construct (which could be expressed, but awkwardly, using the previous syntax), which overcomes some of the problems of "flat" guards (conditionals on the left hand side of rules being only system primitives, not user-defined expressions). The if-then-else takes the form `<Decl>?<Cond>:<Statement>:<Statement>`. Here `<Decl>` is an optional statement of local variables, taking the same form as previous local declarations, `<Cond>` is an expression which evaluates to `true` or `false`, and `<Statement>` has the same structure as the rhs of a rule (including the possibility of further embedding). `<Cond>` is part of the statement preceding the if-then-else, thus having variable values as input into the procedure, but the `<Statement>`s describe the embedded procedure, and thus have variable values as changed by any assignments. The embedded procedure has an extra argument, the result of the expression `<Cond>`, and just two non-recursive rules given by the `<Statement>`s, the first taken if `<Cond>` evaluates to `true`, the second if it evaluates to `false`. As an example, here is a version of list filtering, using the linked list form of lists given above, and the filtering function passed in as an argument:

```
#filter(Func,InList)→OutList
{ || ? InList.isempty
        : empty()→OutList;
        : InList.head→H.tail→T,<(H,T)
          ? Func H
              : filter(Func,T).cons(H)→OutList;
              : filter(Func,T)→OutList
}
```

A more concise form involving implicit recursion and anonymous output is:

```
#filter(Func,InList)~
{ | ? InList.isempty
      :| <=empty();
      :  InList.head→H.tail→InList,<(H)
         ? Func H
             : <=~cons(H);
             :
}
```

where `<=` is a new construct introduced to assign a value to the anonymous object output, equivalent to Actors "become". Without the embedded procedures, this would be:

```
#filter(Func,InList)~ ==
          <=filter1(InList.Isempty,Func,InList);

#filter1(t,Func,InList)→OutList
{
 t=true   ||    empty()→OutList;
 t=false  ||    InList.head→H,
                filter2(Func H,H,Func,InList.tail)→OutList
}
```

```
#filter2(t,H,Func,InList)→OutList
{
 t=true   ||  filter(Func,InList).cons(H)→OutList;
 t=false  ||  filter(Func,InList)→OutList
}
```

A further control structure gives sequencing. A statement may take the form +`<Decl><Statement>;<Statement>`, with `<Decl>` again an optional declaration of further local variables taken from the first of the `<Statement>`s added to the arguments of the embedded procedure, and the second `<Statement>` being the sole rule, entered without condition, of this embedded procedure. The recursive call from this second statement will return to the calling procedure. Again, the `<Statement>`s may contain further embedding. Note the "sequencing" here refers to values of variables: the first statement may change a variable value, the second will make use of the changed value and may change it again. It does not necessarily apply to execution order of every component of the statements, components of the first statement could indeed be suspended waiting for values to be returned from the second.

## 10. Conclusion

A language with a simply rule-based operational model, originally inspired by the concurrent logic languages, has been constructed by adding derived forms to that model. Following from concurrent logic programming, the language uses single-assignment variables, but it imposes a type discipline which ensures any variable has exactly one writer. A variable in a particular position has either input type or output type, and it also has either object type or simple type (where a simple type variable can be bound to a constant or a tuple containing only input simple variables).

We have described briefly, using examples, a variety of the features which are included in Aldwych. The flexibility of the underlying model is clear from this. Currently the language is at an experimental stage. We welcome further discussion in order to settle on a suitable set of features to be included in the first distribution of the language. We would prefer to think of Aldwych as it currently exists as akin to Landin's ISWIM [16], a basis for a family of languages rather than a language in its own right. However, a working version which includes all the features mentioned in this paper is in existence, and has been used to construct several medium-sized programs. The productivity improvement when using it and compared to constructing the same program directly in a concurrent logic language is large.